\documentclass{elsart}
\usepackage{epsfig}
\newcounter{bla}

\begin{document}
\begin{frontmatter}

\title{Atmospheric MUons from PArametric formulas: a fast GEnerator
for neutrino telescopes (MUPAGE)}

\author[a]{G. Carminati},
\author[a]{M. Bazzotti},
\author[a]{A. Margiotta},
\author[a]{M. Spurio\thanksref{author}}

\thanks[author]{Corresponding author}

\address[a]{ Department of Physics, University of Bologna and INFN
Sezione di Bologna, Viale Berti Pichat 6/2, I-40127 Bologna, Italy }

\begin{abstract}
Neutrino telescopes will open, in the next years, new opportunities in observational high energy astrophysics. In these detectors, atmospheric muons from primary cosmic ray interactions in the atmosphere play an important role, because they provide the most abundant source of events for calibration and test. On the other side, they represent the major background source. 

In this paper a fast Monte Carlo generator (called MUPAGE) of bundles of atmospheric muons for underwater/ice neutrino telescopes is presented. MUPAGE is based on parametric formulas \cite{paper} obtained from a full Monte Carlo simulation of cosmic ray showers generating muons in bundle, which are propagated down to 5 km w.e. It produces the event kinematics on the surface of a user-defined cylinder, surrounding the virtual detector.
The multiplicity of the muons in the bundle, the muon lateral distribution and energy spectrum are simulated according to a specific model of primary cosmic ray flux, with constraints from measurements of the muon flux with underground experiments. 
As an example of application,  the result of the generation of events on a cylindrical surface of $\sim$ 1.4 km$^2$ at a depth of 2450 m of water  is presented.

\begin{flushleft}

\end{flushleft}


\end{abstract}

\end{frontmatter}


{\bf PROGRAM SUMMARY}

\begin{small}
\noindent

{\em Title of program:} MUPAGE \\
{\em Nature of the physics problem:} Fast simulation of atmospheric muon bundles for underwater/ice neutrino telescopes. \\
{\em Method of solution:} Atmospheric muon events are generated 
according to parametric formulas \cite{paper} giving the flux, the multiplicity, the radial distribution and the energy spectrum. \\
{\em Licensing provisions: none.}                                   \\
{\em Programming language used:} C++ \\
{\em Computers on which the program has been tested:} Pentium M, 2.0 GHz; 
2x Intel Xeon Quad Core, 2.33 GHz \\
{\em Number of processors used:} one                             \\
{\em Operating systems under which the program has been tested:}
Scientific Linux 3.x; Scientific Linux 4.x;  Slackware 12.0.0 \\
{\em Number of bits in a word:} 32 \\
{\em Memory required to execute program with typical data:} 50 MB  \\
{\em Has the code been vectorized or parallelized?} no \\
{\em Number of lines in distributed program:} 1 933\\
{\em Number of bytes in distributed program:} 59 166\\
{\em Distribution format:} tar \\
{\em Library used:} ROOT \cite{root}  \\
{\em Additional comments:} The program requires the ROOT libraries
for the pseudorandom number generator.\\
{\em Restrictions of the program:} Water vertical depth range from
1.5 to 5 km w.e.; zenith angle range from 0 to 85 degrees. 
{\em Classification:} 1.1 Cosmic Rays; 11.3 cascade and shower simulation.\\
  \\

\end{small}

\newpage



\section{Introduction}
Most astrophysical sources are expected to produce neutrinos in
addition to  photons and cosmic rays \cite{rassegna}. Theoretical
predictions for neutrino fluxes indicate that $\sim 1$ km$^3$ scale
apparatus is probably needed. A neutrino
telescope in the South Pole is currently under construction and is taking data
 with a reduced configuration \cite{icecube}. In the Northern 
hemisphere, the European consortium KM3NeT \cite{km3net}, including
the three collaborations, ANTARES \cite{antares}, NEMO \cite{nemo} and NESTOR \cite{nestor}, is working on a design study for a large deep-sea neutrino telescope in the Mediterranean sea.

Although neutrino telescopes are located under large depth of water
or ice, a great number of high energy atmospheric muons can reach
the active volume of the detectors. Atmospheric muons are produced in the decay of charged mesons generated by the
interactions of primary cosmic rays with atmospheric nuclei. They
represent the most abundant signal in a neutrino telescope and they can
be used to calibrate the detector and to check its response
to the passage of charged particles. On the other side, they can
constitute the major background source, mainly because
downward-going muons can incorrectly be reconstructed as
upward-going particles mimicking high energy neutrino interactions.
Muons in bundles seem to be particularly dangerous \cite{icrc_ant}.
A full Monte Carlo (MC) simulation, starting from the generation of atmospheric showers, can accurately reproduce the main features of muons reaching a neutrino
telescope, but it requires a large amount of CPU time.

In this paper an event generator (MUPAGE: MUon GEnerator from PArametric formulas) based on parametric formulas \cite{paper} is presented. The formulas allow to calculate the flux and angular distribution of underwater/ice muon bundles, taking into account the muon multiplicity and the multi-parameter dependent energy spectrum. The range of validity extends from 1.5 km to 5.0 km of water vertical depth, and from $0^\circ $ up to $85^\circ $ for the zenith angle. 
MUPAGE output is an ASCII table containing the kinematics of events at the surface of a $can$. The $can$ is an imaginary cylinder surrounding the active volume of a generic detector. The ASCII table can be used as input in the following steps of a detector-dependent MC simulation, which include production of light in water/ice and simulation of the signal in the detection devices.

This paper is organized as follows. In Section \ref{sec:para}  the main features of the underwater muon flux, described by the parameterizations of \cite{paper}, are reviewed. In Section \ref{sec:program} the MUPAGE structure is
presented, in particular details are given on how to run the
program, the input parameters and the output file format. Section \ref{sec:gene} describes the generation of events on the surface of the  $can$, with the use of a {\it Hit-or-Miss} method. Events can be single muons (Section \ref{sec:singlemu}) or the more complex multiple muons (Section \ref{sec:multimu}). MUPAGE produces events with the same relative weight. For each run the livetime is computed as described in Section \ref{sec:livetime}. Finally, as an example of application, the case of the ANTARES detector is presented in Section \ref{sec:antares}.

\section{Simulation of bundles of atmospheric muons}\label{sec:para}
Atmospheric muon flux in the deep water is at present experimentally poorly known \cite{grieder}. Some parameterisations for the underwater flux and energy spectrum are available in literature \cite{okada,bugaev,klimu}. None of them 
gives the possibility to simulate
two or more muons produced in the same cosmic ray interaction (muon bundles). To overcome this limitation, at a price of huge CPU time requirement, full simulations of atmospheric showers induced by primary cosmic ray interactions are performed by neutrino telescope collaborations \cite{brunner,icecubediff}. These full MCs are generally based on the CORSIKA \cite{corsika} or HEMAS \cite{hemasdpm} packages. They are then 
followed by the propagation of the surviving  muons from sea level to the detector position. 

MUPAGE relies on parametric formulas which describe the flux, the angular distribution and the energy spectrum of muon bundles for vertical  depths $h$ from 1.5 to 5 km w.e. of water or ice.  The muon energy depends on $h$, on the zenith angle $\vartheta$ and, for muon in bundles, on bundle multiplicity $m$ and on the distance $R$ of each muon with respect to the shower axis. 
The parametric formulas were obtained starting from a full MC simulation of 
primary cosmic ray  interactions and shower propagation in the atmosphere using the HEMAS code. HEMAS was preferred since it was largely used and cross-checked with MACRO data, a large underground experiment operating from 1994 to 2000 and located at the INFN Gran Sasso laboratory, at a depth comparable to that of neutrino telescopes. In particular, MACRO measured the flux of muon bundles \cite{macro-comp} and the distribution of distances between muon pairs in a muon bundle (the so-called decoherence distribution \cite{macro-deco}). The input primary cosmic ray spectrum used in MUPAGE was an unpublished MACRO model bounded by the measurements of underground muons. To optimize the full MC simulation, the code was restricted to follow secondary particles with energies $E>$ 500 GeV.
Muons reaching the sea level with energies $E > 500$ GeV were propagated through water down to 5 km using the MUSIC (MUon SImulation Code) program \cite{music}.
The so called {\it prompt muons} (originated from the decay of charmed mesons and other short-lived particles) were not included. They were expected to give a  not negligible  contribution for muon energies from $\sim 10$  TeV to $\sim 10^3$ TeV, depending on the charm production model \cite{sine}.

A MUPAGE event is a bundle of muons with multiplicity $m_c$ on the $can$. Muons in the bundle are assumed to be parallel to the shower axis, and to reach at the same time a plane perpendicular to the shower axis.
The bundle multiplicity, direction and impact point of the shower axis on the \textit{can} surface are generated first. Then, for each muon in the bundle, the distance from the shower axis, the energy and the coordinates of the impact point on the \textit{can} surface are calculated.

\subsection{Parametric formulas for the flux, energy spectrum and radial distance of muons}
\begin{figure}[!t]
\begin{center}
\includegraphics[scale=1.2]{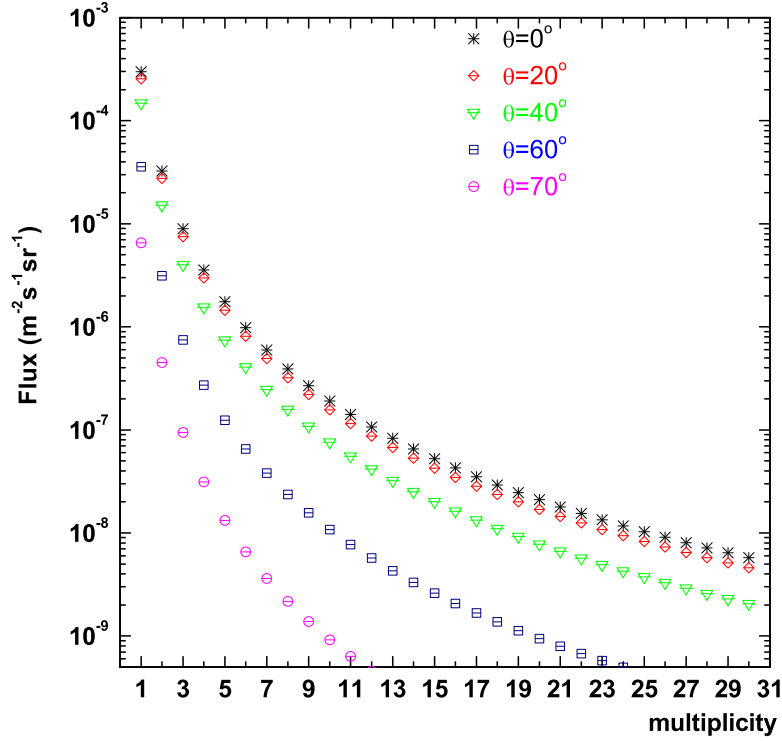}
\caption{\small Flux of muon bundles  with different multiplicity $m$ at the depth of $h=2.5$ km w.e. for five different zenith angles, $\vartheta=0^\circ,20^\circ,40^\circ,60^\circ$ and $70^\circ$ as computed with (\ref{eq:eq1}). } 
\label{fig:multi}
\end{center}
\end{figure}

The flux of bundles with muon multiplicity $m$ at a given vertical depth $h$ and zenith angle $\vartheta$ is parameterized using two free parameters ($K$ and $\nu$) as:
\begin{equation}
\Phi(h,\vartheta,m)= {K(h,\vartheta) \over m^{\nu(h,\vartheta)}} \label{eq:eq1}
\end{equation}
As for the following formulas in this subsection, more details on the functional dependence of the parameters are reported in \cite{paper}.
Fig. \ref{fig:multi} shows the flux of muon bundles  with different multiplicity at the depth $h=2.5$ km w.e. for five different zenith angles, $\vartheta=0^\circ,20^\circ,40^\circ,60^\circ$ and $70^\circ$ obtained using  (\ref{eq:eq1}).

\begin{figure}[!t]
\begin{center}
\includegraphics[scale=1.2]{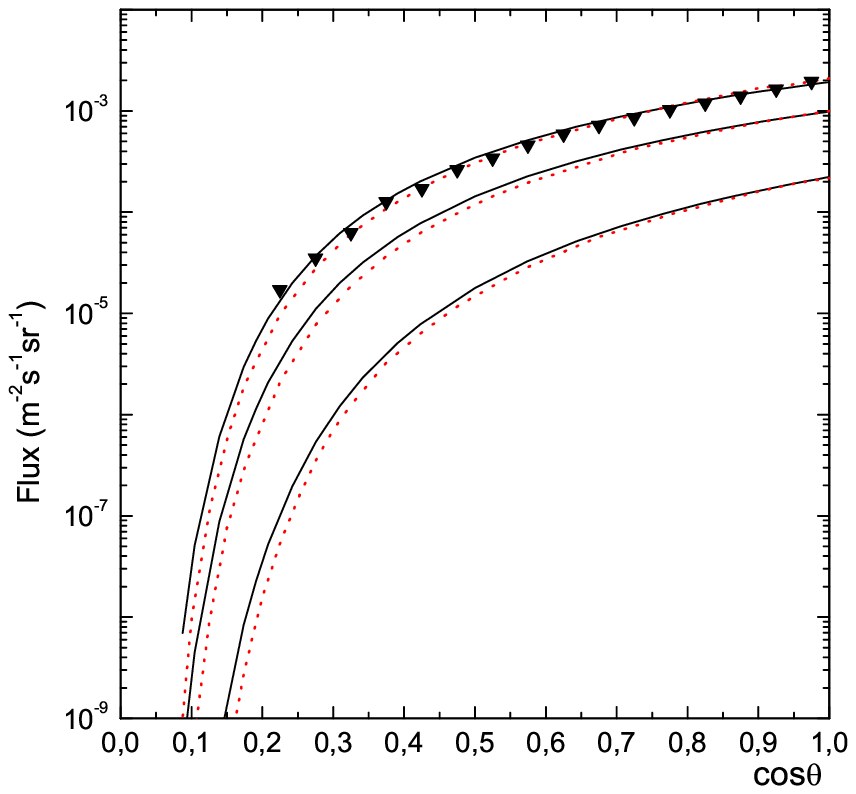}
\caption{\small Muon flux (with energy $E_\mu>20$ GeV) as a function of the cosine of the zenith angle and for three different depths. The full lines are obtained from Eq. (\ref{eq:eq1}), for depths $h=1.64, 2.0$ and $3.0$ km w.e. from top to bottom. The points represent the AMANDA-II measurement \cite{amanda}, at the depths $h=1.64$ km w.e. and with the same muon energy threshold. The (red) dotted lines are from \cite{klimu}, at depth $h=1.61, 2.0$ and $3.0$ km w.e.} 
\label{fig:zenith}
\end{center}
\end{figure}

Fig. \ref{fig:zenith} shows the flux of muons (with energy $E_\mu>20$ GeV) versus the zenith angle  cosine at three different depths. In this case, a bundle of multiplicity $m$ is accounted for $m$ muons. The full lines are obtained with  (\ref{eq:eq1}), for depths $h=1.64, 2.0$ and $3.0$ km w.e. The points represent the AMANDA-II unfolded measurement \cite{amanda}, at the depth $h=1.64$ km w.e. and with the same muon energy threshold. The (red) dotted lines are from \cite{klimu}, at depths $h=1.61, 2.0$ and $3.0$ km w.e. In \cite{klimu}, the bundle multiplicity was not taken into account and the depth-intensity relation of Bugaev et al. \cite{bugaev} was used as input. 
In our case, the depth-intensity relation for the vertical direction (called in literature $I(h,0)$) is computed with (\ref{eq:eq1}) and compared with other parameterizations in Fig. 1 of \cite{paper}. The (normalized) difference between what obtained with (\ref{eq:eq1}) and what reported in \cite{bugaev} is -16\% at 1.5 km w.e. and -5\% at 5.0 km w.e.  The differences at the same depths of (\ref{eq:eq1}) with respect to the Okada parameterization \cite{okada}  are +1\% and -7\%, respectively.

The expected energy distribution of muons, assuming a power-law for the primary beam energy, at a slant depth $X= h/cos\vartheta$ is \cite{lista}:
\begin{equation}
{ dN \over d (log_{10}E_\mu) } = G\cdot E_\mu e^{\beta X (1-\gamma)} [E_\mu + \epsilon (1-e^{-\beta X})]^{- \gamma}    
\label{eq:spectrum} 
\end{equation}
The quantities $\epsilon$ and $\gamma$ are considered as free parameters, while $\beta$ as a constant. $G=G(\gamma,\epsilon)$ represents a normalization factor, in order to get the integrated muon energy spectrum up to $5\times 10^5$ GeV equal to 1. MUPAGE uses (\ref{eq:spectrum}) to extract the muon energy, both for single muon events and for muon bundles.

\begin{itemize}
\item {\bf Single muons (bundles with multiplicity $m=1$)}. The parameter $\gamma$ depends on the vertical depth $h$ only:  $\gamma=\gamma (h)$. The parameter $\epsilon$ depends both on $h$ and on the zenith angle $\vartheta$: 
$\epsilon=\epsilon(h,\vartheta)$. See Section \ref{sec:singlemu}.

\item {\bf Multiple muons (bundles with multiplicity $m>1$)}. Muons produced in the decay of secondary mesons follow the energy distribution of the parent mesons. As a consequence, the most energetic muons are expected to be closer to the axis shower. To sample the energy of a muon in a bundle, its radial distance $R$ from the shower axis is taken into account.
The muon radial distance distribution is described as:
\begin{equation}
{dN \over dR} = C { R\over (R+R_0)^\alpha }
\label{eq:radial} 
\end{equation}
where the parameter $R_0$ depends on the vertical depth $h$, on the bundle multiplicity $m$ and on the zenith angle $\vartheta$: $R_0=R_0(h,m,\vartheta)$; $\alpha$ depends on $h$ and $m$: $\alpha=\alpha(h,m)$. Finally, $C=C(R_0,\alpha)$ is a normalization factor.

For each muon in the bundle, the energy is extracted with a probability given by the distribution (\ref{eq:spectrum}). In this case, the parameter $\gamma$ depends on the vertical depth $h$, on the bundle multiplicity $m$ and on the radial distance $R$: $\gamma=\gamma(h,m,R)$. The parameter $\epsilon$ depends on $h$, on the zenith angle $\vartheta$, and on $R$: 
$\epsilon=\epsilon(h,\vartheta,R)$. See Section \ref{sec:multimu}.
\end{itemize}

Eqs. (\ref{eq:eq1}), (\ref{eq:spectrum}) and (\ref{eq:radial}) are implemented in the MUPAGE \texttt{Muons} class as described in the next section. The values of the numerical constants are implemented in the \texttt{Parameters} class.

\section{Program structure}\label{sec:program}
\begin{table}[!h]
\begin{center}
\begin{tabular}{||l|l|l||}
\hline \hline
\textbf{File name} & \textbf{File type} & \textbf{Contents} \\
\hline
\textbf{Makefile} & make file & file to create the executable \\
\textbf{run-mupage.csh} & C-Shell script file & template script  (tcsh) \\
\textbf{run-mupage.sh} & Shell script file & template script (bash) \\
\textbf{README} & ASCII file & instructions on running MUPAGE \\
\textbf{dat} & directory & input parameter files (see Section
\ref{ssec:input})\\
\textbf{evt} & directory &  event output files (see Section \ref{ssec:output})\\
\textbf{inc} & directory & C++ include files for MUPAGE \\
\textbf{livetime} & directory & run livetime (see
Section \ref{sec:livetime})\\
\textbf{src} & directory & MUPAGE C++ source code \\
\hline \hline 
\end{tabular}
\small \caption{Description of the files contained in the $tar$ file
main directory \texttt{mupage/} } 
\label{tab:struct}
\end{center}
\end{table}

MUPAGE code is written in C++ and it has been tested with gcc
version 3.2.x., 3.4.x and 4.1.x. The program requires  ROOT
libraries \cite{root} for the pseudorandom number generator (see
Section \ref{ssec:axis}). Files  from the $tar$ archive are extracted in the main folder \texttt{mupage/}. This folder contains the \texttt{Makefile},
a \texttt{README} ASCII file, two template script files (for
tcsh and for bash) to launch the executable and some
subdirectories as described in Table \ref{tab:struct}. After running \texttt{Makefile}, a  \texttt{Linux/} directory is created, containing the object files; the executable is created in the main directory. 
The prototypes files \texttt{(.hh)} are in the \texttt{inc/} directory and the declaration files \texttt{(.cc)} in the \texttt{src/}  directory. They do not need to be changed by the user:

\begin{tabular}{ll}
\texttt{inc/ConvertingUnits.hh} : & units conversion \\
\texttt{inc/Geometry.hh} : & geometry definition\\

\texttt{inc/Parameters.hh \& src/Parameters.cc}: & class with parameters
from \cite{paper} to \\
& compute the parametric formulas \\
\texttt{inc/Decode.hh \& src/Decode.cc} : & class to decode input parameters \\
\texttt{inc/Muons.hh \& src/Muons.cc} : & class to generate an event \\
\texttt{src/mupage.cc} : & main program \\
\end{tabular}

The simplest way to execute
MUPAGE is the use  of the C-shell script file \texttt{run-mupage.csh} for tcsh, or the Shell script file \texttt{run-mupage.sh} for bash.
In both template scripts the user can modify  random seed,
run number and the number $N_{gen}$ of events to be generated.
It is also possible to define the directories for MUPAGE output files. 
The livetime of the run is evaluated from $N_{gen}$  (see Section \ref{sec:livetime}). The following control cards, defined in the template scripts, can be set as \texttt{-character [name of data]}:

\begin{tabular}{lll}
\textbf{-h} & &  help \\
\textbf{-i} & \textbf{\$run\_id} & run number \\
\textbf{-n} & \textbf{\$num\_events} & number of events to be generated \\
\textbf{-s} & \textbf{\$seed} & random seed (default 0) \\
\textbf{-p} & \textbf{\$INFILE} & full name (including path) of the
input file (see Section \ref{ssec:input})\\
\textbf{-o} & \textbf{\$OUTFILE1} & full name (including path) of
the first output file \\
& \textbf{\$OUTFILE2} & full name (including path) of the second output file \\
& & (for both output files see Section \ref{ssec:output}) \\
\end{tabular}

Card values are provided as arguments to the executable:

\texttt{./mupage.exe -i \$run\_id -n \$num\_events -s \$seed  -p \$INFILE} \\
\texttt{-o \$OUTFILE1 \$OUTFILE2}

\subsection{Description of the input file}\label{ssec:input}

\begin{figure}[!t]
\begin{center}
\vspace{9.0cm}
\includegraphics{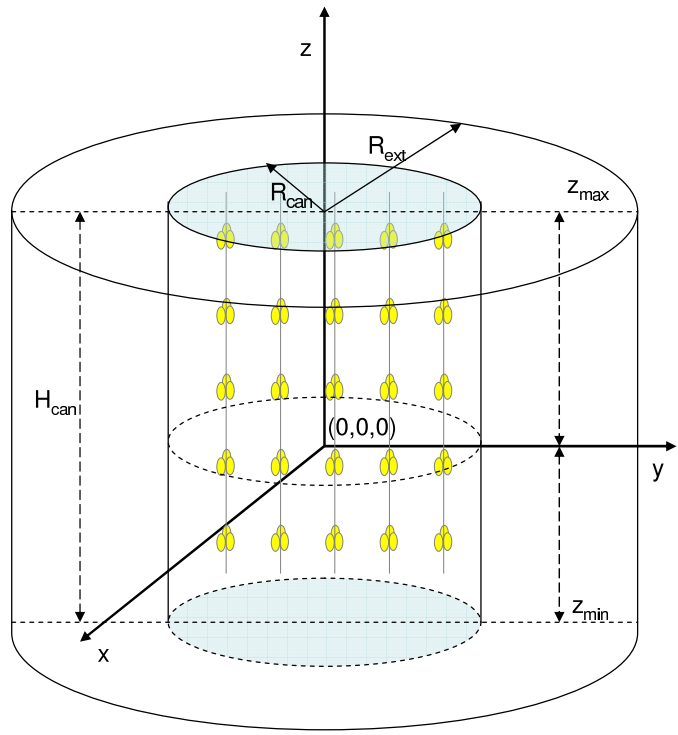}
\caption{\small Sketch of the geometrical meaning of some input parameters. The cylinder surrounding the instrumented volume is the $can$, with radius
$R_{can}$ and height $H_{can}$. The events are generated on an
extended can, with $R_{ext}= R_{can}+ R_{ecr}$.  The origin of coordinate system lies on the cylinder axis, but not necessarily at the center of the
cylinder. The lower disk is at a depth $H_{max}$ with respect to
the sea/ice surface. The origin of the
detector coordinate system lies on the cylinder axis, but it does not
necessarily coincide with the center of the cylinder.} \label{fig:cylinder}
\end{center}
\end{figure}

The generator needs some parameters as input, which are contained in
the file \texttt{parameters.dat} in the directory \texttt{dat/}. Parameters refer to the detector configuration (see Fig.
\ref{fig:cylinder}), to the medium and to the range of 
simulation parameters (multiplicity, zenith angle, muon energy). Name, default value and unit of parameters are described in
\mbox{Table \ref{tab:input}}\footnote{Default values refer to the
ANTARES experiment in the Mediterranean sea, see Section \ref{sec:antares}.}.

\begin{table}[!t]
\begin{tabular}{l|r|l @{}}
\textbf{Parameter} & \textbf{Default value}& \textbf{Description} \\
\hline
\textbf{Hmax} ($H_{max}$) & 2.475 km   & vertical height of the sea/ice level \\
& & with respect to the \textit{can} lower disk \\

\textbf{Zmin} ($Z_{min}$)& -278.151 m & minimum z-coordinate of the \textit{can} w.r.t. \\
              &            & the center of gravity system of detector  \\
\textbf{Zmax} ($Z_{max}$)& 313.971 m  & maximum z-coordinate of the \textit{can} w.r.t.  \\
              &            & the center of gravity system of detector  \\
\textbf{CANr} ($R_{can}$) & 238.611 m  & \textit{can} radius  \\
\textbf{EnlargedCANr} ($R_{ecr}$)& 300.0 m & increase of the \textit{can} radius \\
\textbf{density} & 1.025 g cm$^{-3}$ & mean value of medium density  \\
\textbf{AbsLength} ($\lambda_{abs}$) & 55.0 m& medium absorption length  \\
\textbf{THETAmin} ($\vartheta_{min}$)& $0.0^\circ$ & minimum zenith angle of the bundle \\
\textbf{THETAmax} ($\vartheta_{max}$)& $85.0^\circ$ & maximum zenith angle of the bundle \\
\textbf{Rmin} ($R_{min}$) & 0.0 m& minimum lateral spread of multiple muons  \\
\textbf{Rmax} ($R_{max}$)& 100.0 m & maximum lateral spread of multiple muons  \\
\textbf{Emin} ($E_{min}$)& 0.02 TeV & minimum muon energy at the $can$ level  \\
\textbf{Emax} ($E_{max}$) & 500.0 TeV & maximum muon energy at the $can$ level  \\
\textbf{Ethreshold} ($E_{thr}$) & 0.02 TeV& threshold energy =  $\sum_{i=1}^{m_c} E_{\mu ,i}$ \\
\textbf{MULTmin} ($m_{min}$)& 1 & minimum muon multiplicity \\
\textbf{MULTmax} ($m_{max}$)& 1000 & maximum muon multiplicity \\
\textbf{GEANTid} & 6 & muon ID (default: the GEANT3 id \cite{geant})\\
\textbf{MFactor} & 1.0 & multiplicative factor for special geometries \\
\hline
\end{tabular}
\caption{Description and default values of parameters in the input file
  \texttt{/dat/parameters.dat}.}\label{tab:input}
\end{table}

Note that, in the input file, the \textbf{MULTmax} parameter has a
default value equal to 1000. This is the only parameter affecting
CPU time significantly. To optimize  CPU time, the
event files can be produced with different ranges of
[\textbf{MULTmin}, \textbf{MULTmax}]. In this case, when filling
histograms,  results must be normalized taking into account 
different values of livetime for each file (see Section \ref{sec:antares}) .

\subsection{Description of the output files}\label{ssec:output}

The code provides two output data files: \textbf{\$OUTFILE1} e
\textbf{\$OUTFILE2}. The former is written (by default) in the
directory \texttt{mupage/evt} and contains all information about the
generated events in a formatted ASCII table. An example of output
data file with 1000 generated events (\texttt{run\_01.evt}) is present in the
folder. Each line of the table contains the following information:

\textbf{event\_id mult $track\_id$ $x_{i}\ y_{i}\ z_{i}\ v_{x}\ v_{y}\ v_{z}\
E_{i}\ t_{i}\ \mu\_id$}

where:
\begin{itemize}
\item  \textbf{event\_id} = event number;
\item \textbf{mult} = multiplicity of the muon bundle at the depth where the shower axis hits the \textit{can};
\item for each muon in the bundle and intercepting the \textit{can} $(m_c\le
mult)$:
\item[-]   $track\_id=i$
($i=1,m_c$), muon identifier in the event;
\item[-]  ($x_i$, $y_i$, $z_i$), coordinates of the muon impact
point on the can surface;
\item[-]  ($v_{x}$, $v_{y}$, $v_{z}$), direction cosines of the
muon, coincident with those of the bundle axis;
\item[-] $E_i$ (in GeV), energy of the muon;
\item[-] $t_i$, time delay  of the i-th muon at the \textit{can} surface
with respect to the first muon in the list, $(i=1)$.
$t_i$ can  be either positive or negative;
\item[-] $\mu\_id$, muon particle identification number, for the detector-dependent simulations following the MUPAGE step. By default it is inserted the GEANT3 code $\mu^-$=6. GEANT4 uses the Particle Data Group encoding and $\mu^-$=13 \cite{geant}. By inserting the appropriate ID number for the muon, any other MC can be used. 
\end{itemize}

\begin{figure}[!t]
\begin{center}
\includegraphics[scale=0.65]{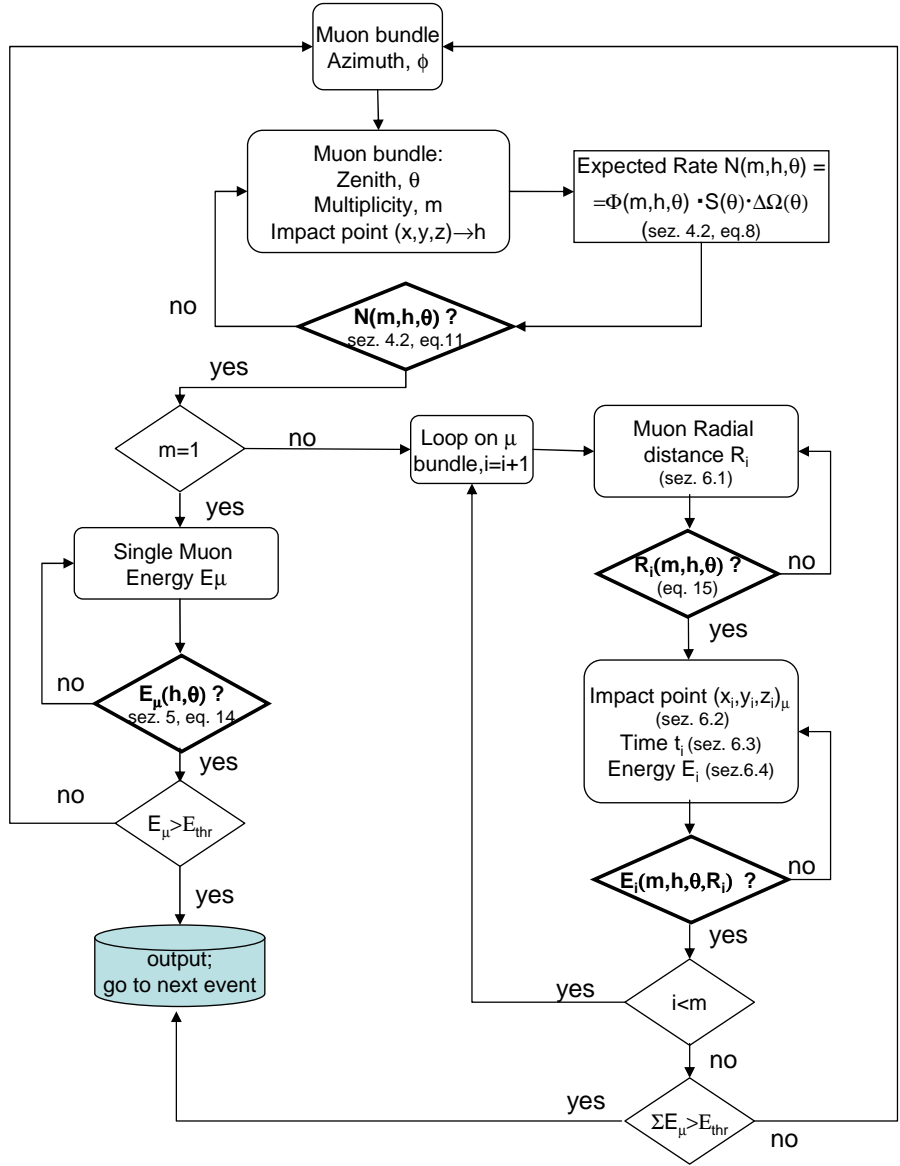}
\caption{\small The flowchart of the MUPAGE event generator. The smooth-angle rectangles indicate the extraction of uniformly  distributed random values. The decisional rhombuses in bold select  values according to formulas reported in Section \ref{sec:para}, with a {\it Hit-or-Miss} method. The procedure is iterated for $N_{gen}$ events. }
\label{fig:flowchart}
\end{center}
\end{figure}

A second output file \textbf{\$OUTFILE2} is written (by default)
in the subdirectory \texttt{mupage/livetime} and contains the
livetime (units: seconds and days) for the simulated run.
The livetime is given with its computed statistical error. The evaluation of livetime is described
in Section \ref{sec:livetime}. All events have the same weight (=1).
The flowchart of the Monte Carlo program is shown in Fig.
\ref{fig:flowchart}; details are described in Sections
\ref{sec:gene}, \ref{sec:singlemu} and \ref{sec:multimu}.

\section{Generation of muon bundles on the \textit{can} surface}
\label{sec:gene}
\subsection{Sampling the bundle direction and impact point on the can surface}
\label{ssec:axis}

Muon bundles at $\sim$ 2 km water equivalent depth can have multiplicity as large as $10^3$, and muons can be hundreds of meters far from the
axis shower. On the other hand, muons travelling few absorption lengths far from the detector have a small probability to give a signal on photomultipliers.
 In order to accept "peripherals" muons in large
bundles, the radius of the generation volume is increased by a
quantity \textbf{EnlargedCANr} ($R_{ecr}$), specified in data cards.
The radius of the generation cylinder is $R_{ext} = R_{can}+R_{ecr} $. 
If $R_{can}$ coincides with the radius of the cylinder surrounding the instrumented volume of the detector, it is recommended to define  $R_{ecr} \simeq 10 \lambda_{abs}$.

\begin{figure}[!t]
\begin{center}
\includegraphics[scale=0.6]{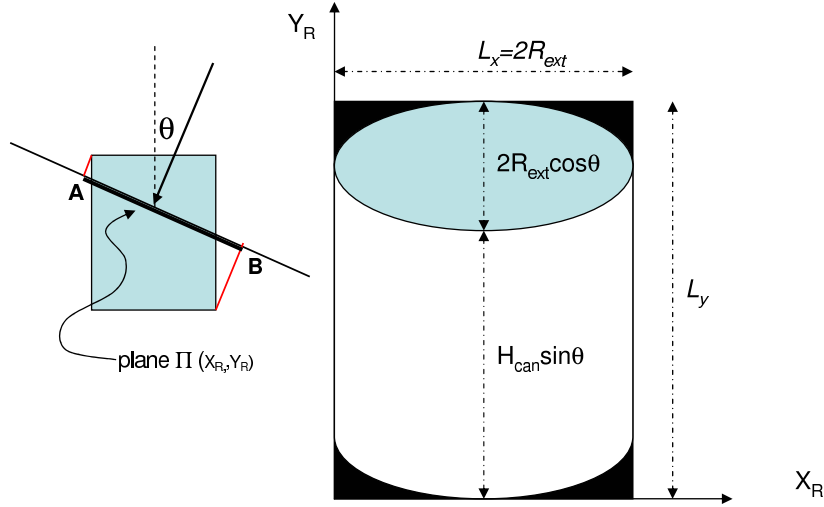}
\caption{\small Left: sketch of the plane $\Pi$ perpendicular to the shower axis. The muon bundle has zenith angle $\vartheta$ with respect to the detector z-axis. The interception point of the shower axis is uniformly distributed on the cylinder projection on the $\Pi$ plane. They will be generated outside the black region; the events in the grey area lie in the upper disk of the cylinder, while
the remaining in the lateral surface. }
\label{fig:intersection_cylinder}
\end{center}
\end{figure}

As a first step, a generic bundle with muon multiplicity $m$* $\in [m_{min}, m_{max}]$, random zenith angle $\vartheta$*$\in [\vartheta_{min},\vartheta_{max}]$ and azimuth angle $\phi$*$ \in [0^\circ , 360^\circ]$ in the detector frame is generated. 
The values that are extracted from uniform distributions, as $m,\vartheta$ and $\phi$ at this step, are denoted with a *. The pseudorandom number generator used in the program is the Mersenne Twister algorithm \cite{MT} and it is included in the ROOT libraries (TRandom3 class) \cite{root}. 
The axis of the bundle with
($\vartheta$*,$\phi$*) intercepts the extended \textit{can} in a random point
of coordinates ($x$*, $y$*, $z$*), which are computed in the following way. The cylinder projected area seen by the bundle  is shown in Fig.
\ref{fig:intersection_cylinder}. The impact point $(X_R,Y_R)$ is a
random point on this plane. Only downward going particles are
generated. It means that points with $z^*=Z_{min}$, on the lower disk of the
\textit{can} are not considered.
$L_x$ and $L_y$ in Fig.
\ref{fig:intersection_cylinder} are defined as $L_x  =  2 R_{ext}$
and $L_y  =  H_{can} \sin\vartheta\textrm{*} + 2 R_{ext}
\cos\vartheta\textrm{*}$. The coordinates $X_R$ and $Y_R$ can assume
the values:
\begin{eqnarray}
- R_{ext} \leq & X_R & \leq R_{ext} \\
- \left( \frac{H_{can}}{2} \sin\vartheta\textrm{*} + R_{ext}
\cos\vartheta\textrm{*} \right) \leq & Y_R & \leq \frac{H_{can}}{2}
\sin\vartheta\textrm{*} + R_{ext} \cos\vartheta\textrm{*}
\end{eqnarray}

The point ($X_R,Y_R$) on the plane perpendicular to the shower
direction can be on the upper disk or on the lateral surface of the \textit{can}. It lies on the upper disk (grey area of Fig. \ref{fig:intersection_cylinder}) if:
\begin{equation}
Y_R > \frac{H_{can}}{2} \sin\vartheta\textrm{*}- R_{ext}
    \cos\vartheta\textrm{*}\quad  \quad
\label{eq:inter_y}
\end{equation}
and
\begin{equation}
\frac{X_R^2}{R_{ext}^2} + \frac{\left[Y_R - (H_{can}/2)
    \sin\vartheta\textrm{*} \right]^2}{(R_{ext}
    \cos\vartheta\textrm{*})^2} \leq 1
\label{eq:inter_upper}
\end{equation}

Returning in 3-D, the  point coordinates are:
 ($x$*, $y$*, $z$*), with $x\textrm{*}
= - X_R $, $y\textrm{*} = \displaystyle
\frac{Y_R}{\cos\vartheta\textrm{*}} - \frac{H_{can}}{2}
\tan\vartheta\textrm{*} $ and $z\textrm{*} = Z_{max}$.

If Eqs. (\ref{eq:inter_y}), (\ref{eq:inter_upper}) are not both
true, the
shower axis hits the lateral surface. In this case, the
intersection point has coordinates 
$x\textrm{*} = R_{ext} \cos\phi'
, y\textrm{*} = R_{ext} \sin\phi' , z\textrm{*} = \displaystyle
\frac{Y_R + \cos\vartheta\textrm{*} \sqrt{R_{ext}^2 - X_R^2}}
{\sin\vartheta\textrm{*}} + \displaystyle \frac{Z_{min}+Z_{max}}{2}
$ where $\phi' = \phi\textrm{*} + \displaystyle \frac{3}{2} \pi -
\arccos \left( - \displaystyle \frac{X_R}{R_{ext}} \right)$.
The points ($x$*, $y$*, $z$*) are distributed uniformly on the \textit{can} surface (with the exclusion of the lower disk).

\subsection{Hit-or-Miss method to sample the impact point}
\label{ssec:hitmiss}

The flux $\Phi(h, \vartheta,m)$, Eq. (\ref{eq:eq1}), decreases with increasing depth $h$, zenith angle $\vartheta$ and muon multiplicity $m$, as shown in Figs. \ref{fig:multi} and \ref{fig:zenith}.
The procedure described in Section \ref{ssec:axis} extracts  uniformly
$h$*, $\vartheta$* and $m$*. A \emph{Hit-or-Miss} method \cite{vonNeumann} is used to reproduce the correct dependence of the number of events
on these variables. For each set of parameters ($h$*, $\vartheta$*,
$m$*), the number of events arriving on the projected area $
S(\vartheta\textrm{*}) $ in a small solid angle $\Delta \Omega
(\vartheta\textrm{*})$ centred around $\vartheta$* is computed:
\begin{equation}
N_{proj}(h\textrm{*}, \vartheta\textrm{*}, m\textrm{*}) =
   \Phi(h\textrm{*}, \vartheta\textrm{*}, m\textrm{*}) \cdot
   S(\vartheta\textrm{*}) \cdot \Delta \Omega (\vartheta\textrm{*})
\end{equation}
where
\begin{equation}
S(\vartheta\textrm{*}) = \pi R_{ext}^2 \cdot \cos\vartheta\textrm{*}
   + 2 R_{ext} H_{can} \cdot \sin\vartheta\textrm{*}
\label{eq:ProjectivArea}
\end{equation}
\begin{equation}
\Delta \Omega (\vartheta\textrm{*}) = 2 \pi
[\cos(\vartheta\textrm{*}-0.5^{\circ})
    - \cos(\vartheta\textrm{*}+0.5^{\circ})]
\label{eq:solidangle}
\end{equation}
A random number $u$ is then generated, with:
\begin{equation}
0 < u < N_{max} \simeq
 \Phi(H_{min},\vartheta_{min}, m_{min}) \cdot S(\vartheta')
\cdot \Delta \Omega (\vartheta')
\label{eq:maxim}
\end{equation}
$ N_{max}$ corresponds to the set of values  $(h,\vartheta,m)$ for which the function $\Phi(h,\vartheta, m) \cdot  S(\vartheta) \cdot \Delta \Omega (\vartheta)$ is maximum. This function has a maximum for the minimum value of the detector depth
$(h=H_{min})$, corresponding to the $can$ upper disk, and for the minimum value of the range of muon multiplicities $(m'=m_{min})$. The maximization in terms of the $\vartheta$ variable is more complex, due to the not trivial
dependence of $\Phi(h,\vartheta, m) \cdot S(\vartheta) \cdot \Delta \Omega (\vartheta)$ on $\vartheta$.

In order to save CPU time, the maximum of 
$\Phi(H_{min},\vartheta, m_{min})~\cdot S(\vartheta)~\cdot~\Delta
\Omega (\vartheta)$ is computed as the product of the maximum of
the functions $\Phi(H_{min},\vartheta, m_{min})$ and $S(\vartheta) \cdot  \Delta \Omega (\vartheta)$. The former has a
maximum in correspondence of  $\vartheta_{min}$. The latter has a
maximum, from (\ref{eq:ProjectivArea}) and (\ref{eq:solidangle}), for zenith angle $\vartheta' = \arctan \left(
\displaystyle \frac{\pi R_{ext}}{2H_{can}} \right)$. Using this
approximation $N_{max}$ is evaluated as in (\ref{eq:maxim}). The
parameter set ($h$*, $\vartheta$* ,$m$*) is accepted if:
\begin{equation}
u < N_{proj}(h\textrm{*}, \vartheta\textrm{*}, m\textrm{*})
\end{equation}
In the following, to simplify the notation, ($h$, $\vartheta$ ,$m$)
will be used and the impact point coordinates become ($x$, $y$,
$z$).

If the \textit{can}  height is
much larger than the disk diameter, $H_{can} \gg R_{ext}$, the approximation used for $ N_{max}$ in (\ref{eq:maxim}) is not valid. It must be stressed that this is NOT the case for present neutrino telescope configurations. However, to correct the procedure, a multiplicative factor in
the input parameters is introduced (default value =1). When the error message appears:
\begin{verbatim}
ERROR! Nmax must be larger than Nproj
\end{verbatim}
and the program stops, the user must change \textbf{MFactor} in
the input data file \texttt(parameters.dat)  to a value larger than 1
(usually \textbf{MFactor} $\sim$  1.2 is enough). As an alternative, the value of $R_{ecr}$ can be increased. 

\section{Single muons}\label{sec:singlemu}

The underwater/ice  flux of atmospheric muons  is dominated (Fig. \ref{fig:multi}) by events reaching the detector with multiplicity $m= 1$, the so-called {\it single muons}.
In this case the muon direction is assumed coincident with the shower axis and the impact point is ($x$, $y$, $z$). The arrival time of the muon on the \textit{can} surface is  $t = 0$. The muon energy $E$
is extracted according to (\ref{eq:spectrum}), whose parameters depend on the vertical depth $h$ of the impact
point on the \textit{can} surface and on the muon zenith angle $\vartheta$.

A value of $\log_{10} E$* is generated randomly between $\log_{10}
E_{min}$ and $\log_{10} E_{max}$ ($E_{min}, E_{max}$ are given in the
data cards). The value $E$* is accepted (or rejected) according to
the {\it Hit-or-Miss} method: a random number $u'$ is generated between 0 and the maximum of (\ref{eq:spectrum}) at depth $h$ and zenith $\vartheta$:  
\begin{equation}
0 < u' < \left( \frac{dN}{d(\log_{10}E_\mu)}
\right)(h,\vartheta;E_\mu^{max})
\end{equation}

The maximum of (\ref{eq:spectrum}) occurs in correspondence of $E_\mu^{max} = { {\epsilon (1-e^{-\beta X}) \over (\gamma - 1)} }$. The value $E$*
is accepted if:
\begin{equation}
u' < \frac{dN}{d(\log_{10}E_\mu)}(h, \vartheta; E^*)
\end{equation}

\section{Multiple muons}\label{sec:multimu}
\subsection{The radial distance of muons with respect to the bundle axis}

For events with muon multiplicity $m> 1$,  the distance $R$ of each muon from
the bundle axis (in a plane perpendicular to the axis) is calculated, according to the radial distribution (\ref{eq:radial}). $R$ depends on  depth $h$, on  bundle multiplicity $m$ and on the zenith angle $\vartheta$. 
It is useful to
define a new reference frame (\emph{Bundle Axis Frame, BAFrame}), where  the $z_{BAF}$-axis coincides with the axis shower, see Fig. \ref{fig:position}. Each muon is located in a point $(X, Y)$ of the plane $\Pi$
perpendicular to the axis shower. The distance of the point $(X,
Y)$ from the origin of the \emph{BAFrame} is called $R_i$ (the
muon radial distance). $R_i$* is sampled randomly between $R_{min}$
and $R_{max}$ (both values from data cards). The {\it Hit-or-Miss}
method is used to accept (or reject) the value $R_i$*. A random
number $u''$ is generated between 0 and the maximum of the lateral
distribution function $(dN/dR)$ at the given $h$, $\vartheta$ and $m$. $R_i$* is accepted if:
\begin{equation}
u'' < \frac{dN}{dR}(h, \vartheta, m, R^*_i)
\end{equation}

The coordinates in the \emph{BAFrame} are computed from the selected $R_i$ as  $X  =  R_i \cdot \cos\beta$ and $Y  = R_i
\cdot \sin\beta$, where $\beta$ is a random number between 0 and
$2\pi$.

\subsection{Coordinates of the multiple muons on the can surface}
\begin{figure}[htb]
\begin{center}
\includegraphics[scale=0.65]{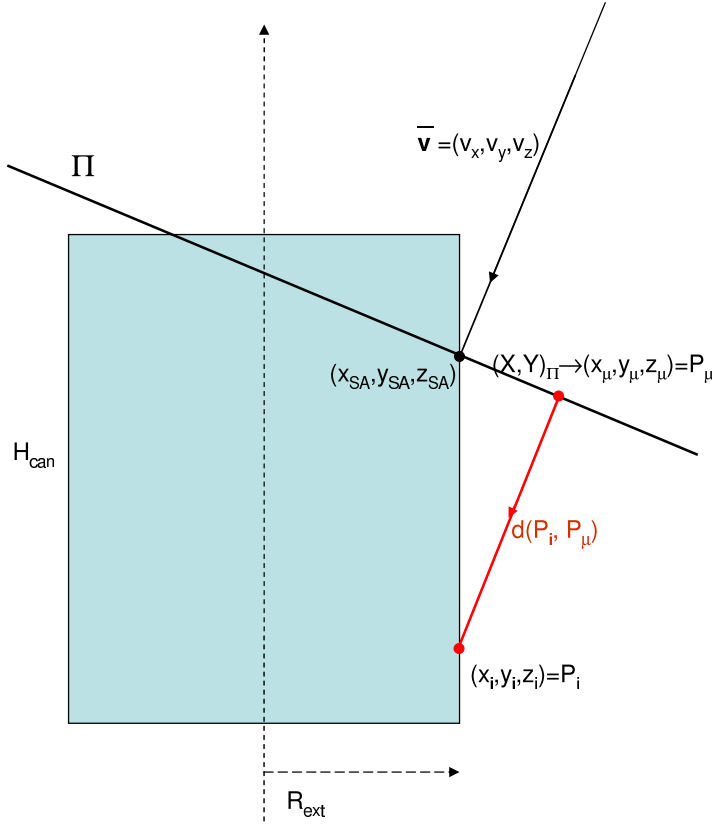}
\caption{\small Lateral view of the extended \textit{can} (the rectangle). The shower axis has direction $(v_x, v_y, v_z)$ and
intercepts the detector in the point $(x_{SA}, y_{SA}, z_{SA})$. $m$
muons are generated on the plane $\Pi$ perpendicular to the shower axis and $m_c\le m$ intercept the $can$. A muon has coordinates $(X, Y)$ in
the \emph{BAFrame} and $(x_\mu, y_\mu, z_\mu)$ in the laboratory
frame. The point $(x_i,y_i,z_i)$ is the projection of the point
$(x_\mu, y_\mu, z_\mu)$ along the $\mu$ direction on the \textit{can}. The time delay  of each muon is evaluated from the distance
between the points $(x_i,y_i,z_i)$ and $(x_\mu, y_\mu, z_\mu)$.}
\label{fig:position}
\end{center}
\end{figure}

The shower axis intercepts the \textit{can} in the impact  point (computed in Section 4.2) with coordinates ($x$, $y$, $z$)=$(x_{SA}, y_{SA}, z_{SA})$.
Then, referring to Fig. \ref{fig:position}, for each muon in the bundle:
\begin{itemize}
\item[-] $(X, Y)$ = coordinates of the muon in the
  \emph{BAFrame};
\item[-] $(x_\mu, y_\mu, z_\mu)$ = coordinates of the muon in the laboratory frame;
\item[-] $(v_x, v_y, v_z) = (\sin\vartheta \cos\phi, \  \sin\vartheta \sin\phi, \ \cos\vartheta)$ = direction of the
    muon in the laboratory frame;
\item[-] $(x_i,y_i,z_i)$ = projection of the point $(x_\mu, y_\mu,
    z_\mu)$ along the shower direction on the \textit{can} surface.
\end{itemize}

When the point $(X, Y)$ in the \emph{BAFrame} is known, the
point $(x_\mu, y_\mu, z_\mu)$ in the laboratory frame can be computed
using a general matrix $\mathbf{A}$ resulting from the
composition of three rotations \cite{euler}.
In the so-called `X-convention' the rotations are defined by the Euler
angles $(\Phi,\Theta,\Psi)$, where the first rotation is by an angle
$\Phi$ around the z-axis, the second one is by an angle $\Theta \in
[0, \pi]$ around the x-axis and the third one is by an angle $\Psi$ around 
the z-axis (again).
There is a univocal relationship between the three Euler angles
and the zenith $(\vartheta)$ and azimuth $(\phi)$ angles:

$\Phi   =  - \pi/2$, $\Theta  =  \vartheta $ and
$\Psi   =  \phi + \pi/2 $.

The transformation of the point $(X, Y)$ in the
\emph{BAFrame} into the point $(x_\mu, y_\mu, z_\mu)$ in the
laboratory frame is defined as:
\begin{eqnarray}
\left(
  \begin{array}{c}
    x_\mu \\
    y_\mu \\
    z_\mu
  \end{array}
\right) = \mathbf{A} \left(
  \begin{array}{c}
    X \\
    Y \\
    0
  \end{array}
\right)
\end{eqnarray}
The coordinates of the impact point of each muon on
the \textit{can} are obtained using the projection of each point
$(x_\mu,y_\mu,z_\mu)$ along the direction $(v_x, v_y, v_z)$. This is
done using the straight line defined by the three parametric
equations $(x_i, y_i, z_i) = (x_\mu,y_\mu,z_\mu) +k(v_x,v_y,v_z)$,
where $ k = (Z_{max} - z_\mu)/{v_z}$.
The impact point of the i-th muon in the bundle is on the upper disk
of the \textit{can} if the straight line intercepts the plane $z = Z_{max}$
with $x_i^2 + y_i^2 \leq R_{ext}^2$. In this case, the coordinates
are:
\begin{equation}
(x_i, y_i, z_i) = (x_\mu + k v_x, \, y_\mu + k v_y, Z_{max})
\end{equation}
If $x_i^2 + y_i^2 > R_{ext}^2$, the impact point $(x_i, y_i, z_i)$
lies on the \textit{can} lateral surface or it does not intercept the \textit{can}
at all.

The intersection of the straight line with the lateral  surface of the \textit{can} 
(defined by equation \mbox{$x^2 + y^2 = R_{ext}^2$})
gives a second degree equation $a \Lambda^2 + 2 b\Lambda + c = 0$,
with $a  =  v_x^2 + v_y^2$, $b  =  v_x x_\mu + v_y y_\mu$ and $c  =
x_\mu^2 + y_\mu^2 - R_{ext}^2$. The solutions are $\Lambda_{\pm} =
\displaystyle \frac{- b \pm \sqrt\Delta}{a}$, with $\Delta = {b^2 -
ac}$. If $\Delta < 0$ the i-th muon does not intercept the \textit{can}. For $\Delta \geq 0$ the two possible impact points are:
\begin{eqnarray}
(x_i, y_i, z_i) = (x_\mu + \Lambda_+ v_x, \, y_\mu + \Lambda_+ v_y, \, z_\mu +
\Lambda_+ v_z) \\
(x_i, y_i, z_i) = (x_\mu + \Lambda_- v_x, \, y_\mu + \Lambda_- v_y, \, z_\mu +
\Lambda_- v_z)
\end{eqnarray}

As atmospheric muons are downward going, the solution with the
larger value $z_i$ is chosen. If $z_i < Z_{min}$ the i-th muon does
not intercept the \textit{can}.
The number of muons intercepting the \textit{can} surface determines the
bundle multiplicity $m_c\le m$ at the \textit{can}.

\subsection{Arrival time of the muons in the bundle}

All muons in the bundle are assumed to arrive at the same time on
the plane $\Pi$ perpendicular to the shower axis. In general,
each muon reaches the \textit{can} surface at a different time.
The distance between the impact point of the i-th muon $P_i(x_i,
y_i, z_i)$ and the coordinates of that muon on the plane $\Pi$ in
the laboratory frame $P_\mu(x_\mu,y_\mu,z_\mu)$ is:
\begin{equation}
d(P_i, P_\mu) = sign \cdot \sqrt{(x_i - x_\mu)^2 + (y_i - y_\mu)^2 +
(z_i - z_\mu)^2} \label{eq:distance}
\end{equation}

The arrival time  of the first muon in the list ($i =1$) on the \textit{can} surface is taken as $t_1 = 0$. All the remaining muons,
labelled with $i = 2,. . .,m_c$, can intercept the \textit{can} earlier $(t_i< 0)$ or later $(t_i > 0)$. The relative time is computed from the
distance $d(P_i, P_\mu)$ defined by (\ref{eq:distance}). A distance is an
always positive quantity, but the evaluation of the relative delay between
muons in the bundle requires the definition of the sign in (\ref{eq:distance}). Referring to Fig. \ref{fig:position}, if $z_\mu <
z_i$ the distance is assumed positive ($sign=1$), otherwise it is
negative ($sign=-1$). The delay  $t_i$ of the i-th muon with
respect to the first one is:
\begin{equation}
t_i = \frac{d(P_i, P_\mu) - d(P_1, P_{\mu1})}{c}
\end{equation}

where $c=2.99 \times 10^8$ m/s is the light
velocity. Since the distances can be either positive or negative,
also $t_i$ (in ns) will assume either positive or negative values.

\subsection{Muon energy for multimuon events }
The last step is the choice of the energy of each muon in a
multimuon bundle according to the energy distribution (\ref{eq:radial}). The muon energy extracted from this distribution depends
on vertical depth, zenith angle, on the multiplicity of the shower and on the radial distance of the muon
from the shower axis. The steps described in Section \ref{sec:singlemu}
for the energy of single muons are repeated for the evaluation of the  energy of each muon in the   bundle.


\section{Livetime of the simulation}\label{sec:livetime}

MUPAGE computes automatically the detector livetime corresponding to the number of generated events $N_{gen}$ on the $can$ surface.
The number of simulated events $N(\Delta \Omega_i)$ in a small
solid angle $\Delta \Omega_i = 2 \pi (\cos\vartheta_{1i} -
\cos\vartheta_{2i})$, with multiplicity $m = m_{min}$ and with shower axis
intercepting the \textit{can} upper disk, are evaluated in 33
bins\footnote{The first bin is $0^o < \vartheta < 10^o$, then 30 bins of 2 degrees are considered. The last two bins are
 $70^o < \vartheta < 76^o$ and $76^o < \vartheta < 85^o$, respectively. The bin size was chosen in order to have a constant or at least an adequate statistical sample in each bin.}. The  expected rate of
muon events with multiplicity $ m_{min}$ on
the \textit{can} upper disk with area $\pi R_{ext}^2$ at the depth $H_{min}$, and in the  solid angle $\Delta \Omega_i$ is:

\begin{equation}
\dot{N}_{MC}( \Delta \Omega_i)={\Phi(H_{min},\vartheta_i , m_{min})
    \cdot S \cdot \Delta \Omega_i} \quad (s^{-1})
\end{equation}

where $\vartheta_i=(\vartheta_{1i}+\vartheta_{2i})/2$, and $S = \pi
R_{ext}^2 \cos\vartheta_i \ (m^2)$ is the projected area of the
upper disk. The equivalent livetime for each bin is:

\begin{equation}
T(\Delta \Omega_i) =N(\Delta \Omega_i )/\dot{N}_{MC}(\Delta
\Omega_i) \quad (s) \label{eq:livetime}
\end{equation}

The livetime with its statistical error is computed as the weighted average of the 33 values
of $ T(\Delta \Omega_i)$ (which have the same value, within statistical errors), and written in the \textbf{\$OUTFILE2} file.


\section{An example of application: the case of ANTARES}\label{sec:antares}

The ANTARES (Astronomy with a Neutrino Telescope and Abyss environmental RESearch) collaboration is operating the largest neutrino telescope in the Northern hemisphere in a site 2475 m deep, 40 km off La-Seyne-sur-Mer (France).
The detector (completed on May 2008) consists of an array of 12 lines separated one from each other on the seabed by 60-80 m. The instrumented area
is $\sim$ 0.06 km$^2$ \cite{antares}. The simulation of atmospheric muons is one of the main task for ANTARES as for other neutrino telescopes. A full MC simulation is used, at the cost price of a large CPU consuming time. The ANTARES software tools are described in \cite{brunner}. 

MUPAGE was used to produce a large sample of atmospheric muons.
A data set with a livetime equivalent to one month of real data was generated with input parameters reported
in Table  \ref{tab:input}. 358 files were created, each one smaller than 2 GB, with $10^7$ events/file and  corresponding to $7260\pm 4$ s. The CPU time required to produce a file (on a 2xIntel Xeon Quad Core, 2.33 GHz) was about 1 hour. It is a relatively large amount of CPU, but it is small when compared to other steps of simulation, namely the tracking and the Cherenkov light emission,
which need $\sim$10 times more CPU. 
Each file, after triggering and data conversion to a format equivalent to raw
data, is equivalent to one run of 2.02 h of real
data and it is used to test the detector response.
The rate of generated events for a detector with parameters equal to the default value of Table \ref{tab:input} is 1240 Hz.
This rate holds for muons above the threshold energy (20 GeV) at the surface of the generation cylinder of 1.4 km$^2$ and it is much larger (a factor almost 1000) than the actually observed event rate after applying a realistic trigger algorithm. To be triggered, muons must pass sufficiently close to  the instrumented detector volume to produce hits on a minimum number of optical sensors.
Analysis of the comparisons between the MUPAGE, the full MC simulation and the data is in progress \cite{margiotta}. 

A much larger data set of atmospheric muons is needed for 
background study of the high energy neutrino ($E_\nu >100$ TeV) diffuse flux \cite{icecubediff,antadif}. In this case, it is not necessary to simulate the bulk of lower energy muon bundles.  
A data set equivalent to one year was generated, using an optimized choice of multiplicity ranges, and a very conservative cut on the energy threshold $E_{thr} = \sum_{i=1}^{m_c} E_{\mu ,i} >$ 3 TeV (set by the parameter
$\mathbf{Ethreshold}$, see Table \ref{tab:input}).
With $E_{thr} >$ 3 TeV, the rate of generated events on the default detector cylinder reduces to 27 Hz. The event multiplicity was divided in 6 sub ranges: $m=1$ (4.7
Hz); $m=2$ (4.4 Hz); $m=3$ (3.1 Hz); $m=4\div 10$ (10.4 Hz);
$m=11\div 100$ (4.2 Hz) and $m=101\div 1000$ (0.04 Hz). 
Each generated file with $m=1,2,3,[4\div 10]$ and $[11\div 100]$ contains $10^6$ events (in order to have size $<$ 2 GB). It respectively corresponding to 59, 63, 90, 29 and 62 hours of livetime. $5\times10^4$ events/file were generated for $m=[101\div 1000]$ 
corresponding to a livetime of 350 hours/file. 
The total number of generated files is 852. The CPU time required  to generate each file ranges from 10 to 30 minutes using the same processor quoted above (it increases with the value of the minimum multiplicity). The total
CPU time needed for the MUPAGE simulation of this data set was 232 hours.

\section{Conclusions}
In this paper, a fast generator (MUPAGE) of the kinematics of atmospheric muon bundles is presented. As input, it uses parametric formulas for the flux of single and multiple muons valid in the range $1.5 \le h \le 5.0$  km w.e. and $\vartheta \le 85^\circ$. The energy spectrum of single and multiple muons is also  simulated, taking into account the dependence of the muon energy on the shower multiplicity and on the distance of the muon from the shower axis. 
The generator represents an useful tool for underwater/ice neutrino telescope to produce large amount of simulated data. 
As an example, the generation rate of atmospheric muon bundles on a cylinder with area of 1.4 km$^2$ laying  at a depth of 2475 m and with total energy larger that 3 TeV is 27 Hz. The $8.5\times 10^8$ events corresponding to one year of data, were produced with 232 hours of CPU on 2.33 GHz processor.

{\bf Acknowledgements}
This work was motivated through studies performed within the ANTARES and NEMO Collaborations. We would gratefully acknowledge the help and the discussions with Tommaso Chiarusi, Giuseppe Codispoti, Irene Madore, Maximiliano Sioli  and Andrea Sottoriva.




\vskip -1.0cm
\begin{thebibliography}{99}

\bibitem{paper}Y. Becherini, A. Margiotta, M. Sioli and M. Spurio, Astrop. Phys.{\bf 25} (2006), 1.

\bibitem{rassegna} J. Learned ,K. Mannheim, Ann.Rev.Nucl. Part. Sci {\bf 50} (2000) 679; F. Halzen, D. Hopper, Rev. Prog. Phys. {\bf 65} (2002) 1025.

\bibitem{icecube} The IceCube Collaboration: contributions to the
30th International Cosmic Ray Conference (ICRC 2007).
arXiv:0711.0353 [astro-ph]. See also: http://icecube.wisc.edu/ 

\bibitem{km3net} A. Kappes, for the KM3NeT Consortium. {\it KM3NeT - A Next
Generation Neutrino Telescope in the Mediterranean Sea}. Contribution
to the 30th International Cosmic Ray Conference (ICRC 2007).
arXiv:0711.0563 [astro-ph]. 

\bibitem{antares} M. Spurio, on the behalf of the ANTARES Coll. {\it  Antares: Towards a Large Underwater Neutrino Experiment}. Contribution to the Rencontres de Physique, La Thuile, 2008.  arXiv:0805.1191. See also: http://antares.in2p3.fr 

\bibitem{nemo} E. Migneco et al. (the NEMO Collaboration). Nucl. Instr. Methods {\bf A567 } (2006) 444. See also: http://nemoweb.lns.infn.it/ 

\bibitem{nestor} E.G. Anassontzis et al. ( the NESTOR Collaboration). Nucl. Instr. Methods {\bf A567 } (2006) 538. See also: http://www.nestor.org.gr 


\bibitem{icrc_ant} S. Cecchini et al. (for the ANTARES Coll.),
Atmospheric muon background in the ANTARES detector, 29th ICRC,
Pune, India.

\bibitem{grieder} P.K.F. Grieder, Cosmic Rays at Earth, Elsevier, Amsterdam, 2001.

\bibitem{okada}  A. Okada, Astrop. Phys. {\bf 2} (1994) 393.

\bibitem{bugaev} E.V. Bugaev et al., Phys. Rev.  {\bf D58} (1998) 054001.

\bibitem{klimu}  S.I. Klimushin et al., Phys. Rev.  {\bf D64} (2001) 014016.

\bibitem{brunner} J. Brunner. {\it Antares simulation tools}.
Proceedings of the VLVnT Workshop, October 5-8, 2003, Amsterdam.
http://www.vlvnt.nl/

\bibitem{icecubediff} A. Achterberg et al, Phys. Rev. {\bf D76} (2007) 042008.
 
\bibitem{corsika} J.N. Capdevielle { et al.}, 
{\em the Karlsruhe extensive air shower simulation code CORSIKA}, 
KFK Report (1992) 4998; FZKA (1998) 6019.

\bibitem{hemasdpm} G.~Battistoni { et al.}, Astrop. Phys. {\bf 3}
(1995) 157.

\bibitem{macro-comp} M.~Ambrosio { et al.} (MACRO Collaboration), 
Phys. Rev. {\bf D56} (1997) 1407; Phys. Rev. {\bf D56} (1997) 1418.

\bibitem{macro-deco} M. Ambrosio { et al.} (MACRO Collaboration), 
Phys. Rev. {\bf D60} (1999) 032001.

\bibitem{music} P. Antonioli et al., Astrop. Phys. 7(1997) 357; arXiv: hep-ph/9705408.

\bibitem{sine} T.S. Sinegevskays, S.I. Sinegovsky,  Phys. Rev. {\bf D63} (2001) 096004.

\bibitem{amanda} P. Desiati, K. Bland et al. (AMANDA Collaboration). {\it Response of AMANDA-II to Cosmic Ray Muons}. 28th ICRC, Tsukuba, Japan, HE 2.3, 1373-1376

\bibitem{lista} P. Lipari, T. Stanev, Phys. Rev.  {\bf D44} (1991) 3543.

\bibitem{geant} (GEANT3) GEANT program manual, CERN program library long writeup W5013,CERN (1993) ; (GEANT4) http://geant4.web.cern.ch/geant4/

\bibitem{vonNeumann} J. von Neumann, Various techniques used in
connection with random digits, Monte Carlo Method, in A.S.
Householder, G.E. Forsythe, and H.H. Germond, eds., National Bureau
of Standards Applied Mathematics Series, N. 12 (Washington, D.C.:
U.S. Government Printing Office, 1951): 36-38.

\bibitem{MT} M. Matsumoto and T. Nishimura, ACM Transactions on Modeling and Computer Simulation,
Vol.8 No.1 (1998) 3.

\bibitem{root} http://root.cern.ch

\bibitem{euler} H. Goldstein, "The Euler Angles" and "Euler Angles in
Alternate Conventions." §4-4 and Appendix B in Classical Mechanics,
2nd ed. Reading, MA: Addison-Wesley, pp. 143-148 and 606-610, 1980.

\bibitem{antadif} A. Romeyer, R. Bruijn, J. de D. Zornoza 
{\it Muon energy reconstruction in ANTARES and its application to the diffuse neutrino flux}. hep-ex/0308074.  28th ICRC, Tsukuba, Japan, July 2003.

\bibitem{margiotta} A. Margiotta, for the ANTARES Coll. {\it Atmospheric muons in the ANTARES detector}, Proc. of the Int. Workshop Very Large Volume $\nu$ Telescope, VLVnT08 - Toulon, Var, France, 22-24 April 2008. In preparation.

\end{thebibliography}
\end{document}